\documentclass[%
 reprint,
 amsmath,amssymb,
 aps,
]{revtex4-1}

\usepackage{color}
\usepackage{graphicx}
\usepackage{dcolumn}
\usepackage{bm}

\newcommand{\comment}[1]{}

\begin{document}

\preprint{APS/123-QED}

\title{Fluctuation-induced First Order Quantum Phase Transition of $U(1)$ Quantum Spin Liquid in Pyrochlore Quantum Antiferromagnet}

\author{Imam Makhfudz}
\affiliation{%
Department of Physics and Astronomy, Johns Hopkins University, 3400
North Charles Street, Baltimore, Maryland 21218, USA
}%


\begin{abstract}
We show using quantum free energy calculation that the quantum phase
transition between $U(1)$ quantum spin liquid (QSL) and
antiferromagnet (AFM) phases in pyrochlore quantum antiferromagnet
(QAFM) is a first order rather than second order. This change in
order from second to first order is induced by gauge fluctuations,
which are explicitly taken into account at gauge theory level in our
effective low energy theory. We therefore have discovered a
fluctuation-induced first order quantum phase transition in
pyrochlore QAFM. We explicitly derive the quantum free energy
description of this QSL to AFM phase transition and show that it is
a weakly first order phase transition. We also briefly discuss the
experimental relevance of this result.



\end{abstract}

\pacs{Valid PACS appear here}
\maketitle


\section{Introduction}

Quantum spin liquid (QSL), the exotic state with no magnetic order
down to very low temperatures has been the subject of intensive
studies since its very first suggestion by P.W. Anderson on the
existence of resonating valence bond type of such state in
triangular lattice \cite{anderson1} and the idea that QSL may be the
physics behind high $T_c$ cuprates \cite{anderson2}. The search for
quantum spin liquid state has expanded to other systems especially
in frustrated quantum magnets \cite{balentsqsl} where geometric and
quantum fluctuations work together to prevent magnetic ordering and
deliver quantum spin disordered states; the QSL. In spin systems,
theoretical studies suggest the existence of phase with Coulomb type
power law correlations, hence called Coulomb phase \cite{2}\cite{1}
and of $U(1)$ quantum spin liquid which is described by emergent
quantum electrodynamics with emergent photons (gauge field) in the
quantum regime. Such $U(1)$ QSL has been argued to exist in
pyrochlore lattice, a 3-d frustrated lattice spins, in the easy axis
limit \cite{1}.

Quantum phase transition between these phases is fascinating problem
because of the possibility for non-Ginzburg-Landau type of
conventional phase transition \cite{senthil}. At finite temperatures
\cite{13}\cite{14} or at low temperatures near quantum criticality
\cite{9}\cite{10}\cite{11}\cite{12}, in systems which involve
coupling of order parameter to soft modes, such as electromagnetic
field or coupling of one order parameter to the phase fluctuations
of competing order, a second order phase transition can be driven to
first order one by the gauge fluctuations. We will show in this
paper that similar effect but at $T=0$ occurs between QSL and its
neighboring AFM phase in pyrochlore QAFM due to quantum fluctuations
and coupling of order parameter to gauge field.

\section{Model of Pyrochlore QAFM}

Pyrochlore QAFM can be described microscopically by the most general
symmetry-allowed microscopic spin Hamiltonian with spin defined on
pyrochlore lattice site \cite{Onoda}\cite{KRoss}. Considering the
case sufficient for pyrochlore compounds with Kramers doublet and
mapping the spin living at the pyrochlore lattice site to the spin
defined on the link of dual diamond lattice followed by the mapping
\cite{3}
\begin{equation}\label{mapping}
S^{+}_{\textbf{r}\textbf{r}'}=\Phi^{\dag}_{\textbf{r}}s^{+}_{\textbf{r}\textbf{r}'}\Phi_{\textbf{r}'}=
\Phi^{\dag}_{\textbf{r}}e^{iA_{\textbf{r}\textbf{r}'}}\Phi_{\textbf{r}'},S^z_{\textbf{r}\textbf{r}'}=E_{\textbf{r}\textbf{r}'}
\end{equation}
one obtains a Hamiltonian describing bosonic spinons hopping between
the sites of dual bipartite diamond lattice interacting with compact
$U(1)$ gauge field \cite{3}.

\begin{widetext}
\begin{equation}\label{latticegauge}
H=\sum_{r\in I,II}\frac{J_{zz}}{2} Q^2_{\textbf{r}}
-J_{\pm}\{\sum_{\textbf{r}\in
I}\sum_{\mu,\nu\neq\mu}\Phi^{\dag}_{\textbf{r}+\widehat{\textbf{e}}_{\mu}}\Phi_{\textbf{r}+\widehat{\textbf{e}}_{\nu}}s^{-}_{\textbf{r},\textbf{r}+\widehat{\textbf{e}}_{\mu}}s^{+}_{\textbf{r},\textbf{r}+\widehat{\textbf{e}}_{\nu}}+
II\} -J_{z\pm}\{\sum_{\textbf{r}\in
I}\sum_{\mu,\nu\neq\mu}\gamma^*_{\mu\nu}\Phi^{\dag}_{\textbf{r}}\Phi_{\textbf{r}+\widehat{\textbf{e}}_{\nu}}s^z_{\textbf{r},\textbf{r}+\widehat{\textbf{e}}_{\mu}}s^+_
{\textbf{r},\textbf{r}+\widehat{\textbf{e}}_{\nu}}+H.c.+ II\}
\end{equation}
\end{widetext}

where $\widehat{\textbf{e}}_{\mu(\nu)}$ are the local $z$ spin axis
unit vectors at the four corners of a tetrahedron of pyrochlore
lattice \cite{3}, which we will eventually use as the equivalent
substitute for global spin space basis vectors \cite{continua} and
$\gamma_{\mu\nu}$ is element of $4\times 4$ matrix \cite{3}. The
vectors $\widehat{\textbf{e}}_{\mu(\nu)}$ encode the symmetries of
the original microscopic lattice spin model whereas the
$\gamma_{\mu\nu}$ matrix encodes the symmetries of the interaction.
The $Q_{\textbf{r}}$ represents the bosonic spinon number operator
satisfying commutation relation
$[\varphi_\textbf{r},Q_\textbf{r}]=i$ where $\varphi_\textbf{r}$ is
the phase of the bosonic creation operator
$\Phi^{\dag}_\textbf{r}=e^{i\varphi_\textbf{r}}$. Using a 'gauge
mean field theory' (gMFT), Ref. \cite{3} has obtained the phase
diagram of Eq. (\ref{latticegauge}) where $U(1)$ quantum spin liquid
phase exists in narrow region in proximity to neighboring
magnetically ordered phases.

\section{Low Energy Effective Theory of Pyrochlore QAFM}

In this work, we derive the continuum effective low energy theory of
pyrochlore quantum antiferromagnet and investigate the quantum phase
transition of $U(1)$ QSL phase of pyrochlore quantum antiferromagnet
taking into account gauge fluctuations and the interaction between
spinon and photon. The low energy effective field theory is the
first main result of this paper and is given in Minkowski space-time
by \cite{continua},

\begin{widetext}
\begin{equation}\label{effaction1}
S=\int d^4 x [\frac{1}{2J_{zz}}|(i\partial_{t}-e_g
A_0)\Phi_{\textbf{r}}|^2-
\frac{16}{3}J_{\pm}\sum_{\alpha}|(i\partial_{\alpha}-e_g
A_{\alpha})\Phi_{\textbf{r}}|^2
-m|\Phi_{\textbf{r}}|^2-u|\Phi_{\textbf{r}}|^4-\frac{1}{2g^2}\sum_{\alpha\beta}(\partial_{\alpha}A_{\beta}-\partial_{\beta}A_{\alpha})^2+\frac{1}{2g^2}\sum_{\alpha}E^2_{\alpha}]
\end{equation}
\end{widetext}

with spinon gap $m$ and $\alpha,\beta=x,y,z$. The spin exchange
constants $J$'s in the original lattice model Eq.
(\ref{latticegauge}) determine the coefficients of the field theory.
This equation describes complex scalar field coupled to $U(1)$ gauge
field. We have explicitly included the Maxwell term, separated into
its magnetic and electric parts, which describes very well the
physics of pyrochlore QAFM. We have written the field theory with
parameters in Gaussian unit where $g^2=\frac{\mu}{\mu_0}$, unit
lattice spacing $a=1$, speed of photon $v_p=1$ and $\hbar=1$, but we
will revert to physical unit whenever it is necessary.

The field theory is valid low energy description of the pyrochlore
QAFM near the minimum of spinon energy dispersion at $\textbf{k}=0$,
which is the case at mean field level \cite{3} for QSL and AFM
phases. The field theory above takes the form of scalar QED
\cite{psqft} with (emergent) $U(1)$ gauge charge $e_g\equiv Q$
\cite{gaugecharge}. The microscopic lattice model
(\ref{latticegauge}) of charged bosons coupled to compact $U(1)$
gauge field has $U(1)$ gauge invariance representing local gauge
charge conservation and the field theory (\ref{effaction1})
preserves this gauge invariance.

This field theory in its Euclidean space-time form and static mean
field case mimics the quantum($T=0$ version of) free energy
description of BCS superconductor in magnetic field \cite{13} with
boson density corresponding to Cooper pair density and the gauge
field corresponding to the electromagnetic field in such system.
This motivates an analogy between the quantum criticality of
pyrochlore QAFM described by Eq. (\ref{effaction1}) and classical
phase transition in such BCS superconductor under magnetic field. In
this case, the spin exchange $J$'s play the role of energy scale
analogous to temperature $T$. In support of this analogy, it is to
be noted that physically, in pyrochlore QAFM, the scalar potential
is zero $A_0=0$. It is also to be noted that by definition, in both
QSL and AFM phases, the expectation value of electric field $E$ in
Eq. (\ref{effaction1}) vanishes $\langle E\rangle=0$ \cite{3}.

\section{Free Energy Description of $U(1)$ QSL-AFM QPT: Fluctuation-induced First Order QPT}

The QSL to AFM phase transition should be able to be described by an
effective action in terms of bosonic spinon field expectation value
$\langle\Phi\rangle$ (as the order parameter, in the language of
Landau symmetry breaking). To arrive at that, we formally integrate
out the gauge fields from our full action (using both static
spatially uniform (mean-field) solution approximation \cite{13} and
functional integration \cite{7}).

\[
Z=\int
\mathcal{D}\Phi^{*}\int\mathcal{D}\Phi\int\mathcal{D}\lambda\int
\mathcal{D}A e^{-S[\Phi^{*},\Phi,\lambda,A]}
\]
\begin{equation}
=\int \mathcal{D} \Phi^* \int \mathcal{D} \Phi
e^{-S_{eff}[\Phi^*,\Phi]}=\int \mathcal{D} \Phi^* \int \mathcal{D}
\Phi e^{-\frac{F[\Phi^*,\Phi]}{T_{QPT}}}
\end{equation}

Here, $\lambda$ is the Lagrange multiplier that globally imposes the
constraint $|\Phi_{\textbf{r}}|=1$ via a term $\lambda \int d^4 x
(|\Phi_{\textbf{r}}|^2-1)$ contributing to the
$m|\Phi_{\textbf{r}}|^2$ in Eq. (\ref{effaction1}). We obtain the
"free energy" (the quantum version analog of classical thermal free
energy) of bosonic spinon fields
$F[\Phi^*_{\textbf{r}},\Phi_{\textbf{r}}]$ with $J_{\pm}$ playing
the role of energy scale $T_{QPT}$ that tunes the QSL-AFM quantum
phase transition \cite{continua},

\begin{equation}\label{cubicfreeenergy1}
F[\Phi^*_{\textbf{r}},\Phi_{\textbf{r}}]=\int d^3 r
[c_2|\Phi_\textbf{r}|^2-c_3|\Phi_\textbf{r}|^3+c_4|\Phi_\textbf{r}|^4]
\end{equation}

where each $\int_{k}$ is $3$-d momentum integral
$\int\frac{d^3k}{(2\pi)^3}$. The coefficients in physical unit are
$c_2=m+\delta m$ with $m=\lambda-12J_{\pm}$, $\delta
m=\frac{16J_{\pm}a^2e^2_gg^2\mu_0J^c_{\pm}\Lambda}{\pi^2\hbar^2}+\mathcal{O}(e^2_gJ^2_{z\pm})$,
$c_3=\frac{16J_{\pm}a^2e^2_g
g^2\mu_0J^c_{\pm}}{3\pi\hbar^2}\sqrt{\frac{32J_{\pm}a^2e^2_g
g^2\mu_0}{3\hbar^2}}$ and $c_4\equiv
u=u_0+\mathcal{O}(J^2_{z\pm}e^4_g)$
\cite{loopcorrection}\cite{quarticJpmpm}.
$J^c_{\pm}=\frac{\lambda}{12}$ is the critical $J_{\pm}$ at which
$m$ changes sign whereas $\Lambda\sim \frac{1}{a}$ with $a$ is
microscopic lattice spacing \cite{continua}. This free energy is the
second main result of this paper. We note that the coupling to gauge
field generates the crucial cubic term with negative coefficient
which gives rise to first order phase transition with order
parameter $\langle\Phi\rangle$ as we change the coupling $J_{\pm}$
while $J_{z\pm}$ mainly gives rise to corrections to spinon gap
(mass) and especially to quartic term which ensures the stability of
the theory \cite{quarticJpmpm}. Increasing $J_{\pm}$ with other
parameters fixed drives bosonic spinon condensation and this
describes the QSL to AFM quantum phase transition.

The location of QSL-AFM phase transition can be predicted directly
from the free energy Eq. (\ref{cubicfreeenergy1}) which can be shown
to suggest that the phase transition occurs at
$c_2=\frac{c^2_3}{4c_4}$. Physically, the QSL to AFM phase
transition is bosonic spinon condensation that occurs once the
spinon becomes gapless. To lowest order approximation, the free
energy Eq. (\ref{cubicfreeenergy1}) predicts the QSL to AFM
transition to occur at $\lambda\simeq 3J_{\pm}$ \cite{continua}, the
third main result of this paper.

A quantity of interest in a first order phase transition is the size
of that transition. From the free energy Eq.
(\ref{cubicfreeenergy1}), if we define the size of first order
transition as the ratio of the jump $\Delta\langle\Phi\rangle$ to
the magnitude of the order parameter $\Phi_0$ deep inside the
magnetically ordered AFM state $J_{z\pm}\ll J_{\pm}\sim J_{zz}$, we
obtain $\frac{\Delta
\langle\Phi\rangle}{\langle\Phi\rangle_0}=\frac{2}{3}(\frac{J^c_{\pm}}{J_{zz}})^{\frac{3}{2}}\comment{\frac{\langle
A_{\alpha}A_{\beta}\rangle^2|_{AFM}}{\langle
A_{\alpha}A_{\beta}\rangle^2|_{QPT}}$ where $\langle
A_{\alpha}A_{\beta}\rangle|_{QPT(AFM)}$ represents the degree of
gauge field fluctuations at the QPT and inside AFM phase
respectively \cite{continua}. The quantum dynamics of gauge field
can be described by quantum harmonic oscillators with zero point
energy so that gauge fluctuations remain nonzero even at $T=0$ and
in magnetically ordered phase. It is however physically clear that
the gauge field fluctuations are stronger near QPT than in deep AFM
state. As a result, we obtain $\frac{\Delta
\langle\Phi\rangle}{\langle\Phi\rangle_0}}\lesssim 0.06$ (using
$J^c_{\pm}\simeq 0.2J_{zz}$; the location of QSL-AFM phase
transition predicted by gMFT \cite{3}), which suggests a relatively
weak first order phase transition. This smallness in the strength of
first order phase transition is one possible reason for why it is
not captured by mean field theory. This is the final main result of
this paper.

We have therefore shown that the gauge field fluctuations have
driven the mean field second order continuous QSL to AFM quantum
phase transition to weakly first order phase transition. An analogy
with the classical thermal fluctuation-induced first order phase
transitions requires a coupling of an order parameter to gauge field
to drive the phase transition to first order. In our work, such
order parameter is the expectation value of bosonic spinon field
which appears in mean field decomposition of the lattice gauge
theory Eq. (\ref{latticegauge});
$\langle\Phi^{\dag}\Phi\rangle=\langle|\Phi|^2\rangle=|\langle\Phi\rangle|^2+\langle|\Delta\Phi|^2\rangle$.
This mechanism is possible because the spinons carry emergent $U(1)$
gauge charge that couple to emergent electromagnetic gauge field
$\textbf{A}$, that is, it all arises from the $U(1)$ gauge structure
of the theory with complex scalar field coupling to the $U(1)$ gauge
field via the $U(1)$ gauge charge. However, different from
previously known fluctuation-induced first order phase transitions
which occur at finite temperatures or at low temperatures near
quantum critical point, in this paper the phase transition is
between zero temperature quantum ground states and the fluctuations
are quantum fluctuations rather than thermal fluctuations.

\section{Discussion}

In this work, we theoretically investigate quantum phase transition
tuned by changing the spin exchange constants
$J_{zz},J_{\pm},J_{z\pm}$. This is a challenging task experimentally
since given a compound with a measured set of couplings, it
corresponds to merely one point in the theoretical phase diagram.
The quantum phase transition studied here is more feasible for
experimental studies if we can find compound which microscopically
is close to criticality. Pyrochlore compounds near quantum critical
point have been the subject of intense research lately \cite{exp1}.
It is our hope that soon there will be discovered compounds with
microscopic parameters near QSL-AFM transition line. Applying
pressure \cite{broholm}, magnetic field or chemical substitution
(doping) to such compounds as indirect means to tune coupling
constants is expected to drive the compound to cross the QSL-AFM
phase boundary where the quantum criticality predicted in this paper
can be directly verified.

We have treated bosonic spinon and gauge field explicitly and the
interaction between them has been included in the free energy
calculation. Other than these two excitations, there is also
electric monopole which is gapped and plays important role
especially at energies above the gap.\comment{Different from 2-d
case where the magnetic monopoles of compact $U(1)$ gauge theory can
lead to confinement and destroy the spin liquid state, in 3-d the
monopoles are suppressed and so we have stable $U(1)$ QSL. This
observation at the same time justifies our low energy effective
theory derivation where the compactness of the gauge field is
sacrificed.} The presence of fractional excitations in pyrochlore
QAFM is itself an exciting question that has been investigated
experimentally \cite{broholmfraction}. It is a challenge to include
all these excitations and treat the interaction between them fully
field theoretically. Another open problem of interest is to do
similar study on the nature of quantum phase transition between
$U(1)$ QSL and the so-called 'Coulombic' ferromagnet (CFM) phase
\cite{3}. CFM is an interesting phase because it has ferromagnetic
order but with spinon and photon as excitations rather than spin
wave. We find that studying QSL-CFM phase transition using similar
free energy description is a more formidable task and is therefore
an open opportunity for further effort.

In conclusion, in this gauge theory picture, we obtain first order
QSL-AFM quantum phase transition driven by gauge fluctuations
treated at gauge theory level. We conclude that gauge fluctuations
have driven the mean-field second order phase transition to first
order one. We therefore obtain a fluctuation-induced first order
quantum phase transition rather than the standard Ginzburg-Landau
theory's continuous second order. To be more precise, this QSL-AFM
phase transition is predicted to be weakly first order. The
occurrence of this phenomenon reflects the $U(1)$ gauge theory
structure of the Coulomb phases of pyrochlore QAFM, which manifests
an emergent quantum electrodynamics (QED).

\begin{acknowledgments}

We thank Oleg Tchernyshyov, Collin Broholm, Predrag Nikolic, Anirban
Ghosh, Arpit Gupta, Liang Dai, Tom Zorawski, and Jingsheng Li for
helpful discussions.

\end{acknowledgments}

\appendix

\section{Derivation of Low Energy Effective Theory}

We begin with the most general symmetry-allowed nearest neighbor
spin exchange model on pyrochlore lattice
 \cite{Onoda}\cite{KRoss}:

\[
H=\sum_{\langle ij\rangle}\{J_{zz} S^z_i S^z_j - J_{\pm}(S^+_i S^-_j
+S^-_iS^+_j)+J_{\pm\pm}[\gamma_{ij}S^+_iS^+_j
\]
\begin{equation}\label{spinmodel}
+\gamma^*_{ij}S^-_iS^-_j]+J_{z\pm}[S^z_i(\zeta_{ij}S^+_j+\zeta^*_{ij}S^-_j)+i\leftrightarrow
j]\}
\end{equation}

with local cubic basis vectors

\[
\widehat{\textbf{e}_0}=(1,1,1)/\sqrt{3},\widehat{\textbf{e}_1}=(1,-1,-1)/\sqrt{3},
\]
\begin{equation}\label{basisvectors}
\widehat{\textbf{e}_2}=(-1,1,-1)/\sqrt{3},\widehat{\textbf{e}_3}=(-1,-1,1)/\sqrt{3}
\end{equation}

with matrix $\gamma_{\mu\nu}$ given by
\begin{equation}
\gamma_{\mu\nu}= \left( \begin{array}{cccc}
0 & 1 & w & w^2\\
1 & 0 & w^2 & w \\
w & w^2 & 0 & 1\\
w^2 & w & 1 & 0 \end{array} \right)
\end{equation}

where $w=e^{i\frac{2\pi}{3}}$. The 4 unit vectors
$\widehat{\textbf{e}}_{\mu}$ physically point from the center of a
tetrahedron to its four corners. Each defines the local spin $z$
axis of the spin at the corresponding site at the corner. We will
however use these 4 unit vectors as (nonorthogonal) global basis
vectors which are directly related to the $\widehat{\textbf{x}}$,
$\widehat{\textbf{y}}$, and $\widehat{\textbf{z}}$ unit vectors of
global 3-d Cartesian coordinate. Their relation can be written as,

\begin{equation}\label{basisvectorsmapping}
\widehat{\textbf{e}}_{\mu}=\sum_{\alpha=1,2,3}n_{\mu\alpha}\widehat{\textbf{e}}_{\alpha}
\end{equation}

where the coefficient $n_{\mu\alpha}$ is nothing but the
$\alpha^{th}$ element of $\widehat{\textbf{e}}_{\mu}$.

We now follow Ref. \cite{3} to map the spin model with spins defined
at the sites of pyrochlore lattice to spin model with spins defined
at the links of dual diamond lattice and finally to lattice gauge
theory with bosonic spinons defined at the dual diamond lattice
sites and the gauge field at the links, via the mapping

\begin{equation}
S^+_{\textbf{r},\textbf{r}+\textbf{e}_{\mu}}=\Phi^{\dag}_{\textbf{r}}s^+_{\textbf{r},\textbf{r}+\textbf{e}_{\mu}}\Phi_{\textbf{r}+\textbf{e}_{\mu}},
S^z_{\textbf{r},\textbf{r}+\textbf{e}_{\mu}}=s^z_{\textbf{r},\textbf{r}+\textbf{e}_{\mu}}
\end{equation}

The correspondence between pseudospin operators and gauge fields is
given by,

\begin{equation}\label{correspondence}
s^z_{\textbf{r},\textbf{r'}}=E_{\textbf{r},\textbf{r'}},s^{\pm}=e^{i\pm
A_{\textbf{r},\textbf{r'}}}
\end{equation}

where $\eta_{\textbf{r}}=\pm 1$ for diamond sublattice I(II).
Considering the case with $J_{\pm\pm}=0$ \cite{3}, the Hamiltonian
of the lattice gauge theory with bosonic spinons defined on the dual
diamond bipartite lattice becomes

\begin{widetext}
\[
H=\sum_{\textbf{r}\in I, II}\frac{J_{zz}}{2}Q^2_\textbf{r} -
J_{\pm}\{\sum_{\textbf{r}\in
I}\sum_{\mu,\nu\neq\mu}\Phi^{\dag}_{\textbf{r}+\widehat{\textbf{e}}_{\mu}}
\Phi_{\textbf{r}+\widehat{\textbf{e}}_{\nu}}
s^-_{\textbf{r},\textbf{r}+\widehat{\textbf{e}}_{\mu}}s^+_{\textbf{r},\textbf{r}+\widehat{\textbf{e}}_{\nu}}
+ \sum_{\textbf{r}\in
II}\sum_{\mu,\nu\neq\mu}\Phi^{\dag}_{\textbf{r}-\widehat{\textbf{e}}_{\mu}}
\Phi_{\textbf{r}-\widehat{\textbf{e}}_{\nu}}
s^+_{\textbf{r},\textbf{r}-\widehat{\textbf{e}}_{\mu}}s^-_{\textbf{r},\textbf{r}-\widehat{\textbf{e}}_{\nu}}\}
\]
\begin{equation}\label{latticebosonu1}
-J_{z\pm}\{\sum_{\textbf{r}\in
I}\sum_{\mu,\nu\neq\mu}(\gamma^*_{\mu,\nu}\Phi^{\dag}_{\textbf{r}}
\Phi_{\textbf{r}+\widehat{\textbf{e}}_{\nu}}
s^z_{\textbf{r},\textbf{r}+\widehat{\textbf{e}}_{\mu}}s^+_{\textbf{r},\textbf{r}+\widehat{\textbf{e}}_{\nu}}+H.c.)+\sum_{\textbf{r}\in
II}\sum_{\mu,\nu\neq\mu}(\gamma^*_{\mu,\nu}\Phi^{\dag}_{\textbf{r}-\widehat{\textbf{e}}_{\nu}}
\Phi_{\textbf{r}}
s^z_{\textbf{r},\textbf{r}-\widehat{\textbf{e}}_{\mu}}s^+_{\textbf{r},\textbf{r}-\widehat{\textbf{e}}_{\nu}}+H.c.)\}
\end{equation}
\end{widetext}

Despite the rather complicated form, the above Hamiltonian has
$U(1)$ gauge symmetry, i.e. it is invariant with respect to $U(1)$
gauge transformations $\Phi^{\dag}_\textbf{r}\rightarrow
e^{-i\chi(\textbf{r})}\Phi^{\dag}_\textbf{r},
\Phi_\textbf{r}\rightarrow e^{i\chi(\textbf{r})}\Phi_\textbf{r},
A_{\textbf{r},\textbf{r}+\widehat{\textbf{e}}_{\mu}}\rightarrow
A_{\textbf{r},\textbf{r}+\widehat{\textbf{e}}_{\mu}}+\chi(\textbf{r})-\chi(\textbf{r}+\widehat{\textbf{e}}_{\mu})$.
The first term is the Ising term expressed in terms of charge
$Q_{\textbf{r}}$ with integer $Q_{\textbf{r}}\in \mathbb{Z}$ which
basically counts the number of spinons at site $\textbf{r}$ of the
dual diamond lattice.

\begin{equation}
Q_{\textbf{r}}=\eta_{\textbf{r}}\sum_{\mu}
S^z_{\textbf{r},\textbf{r}+\eta_{\textbf{r}}\widehat{\textbf{e}}_{\mu}}
\end{equation}

In fact, $Q_{\textbf{r}}$ is the analog of momentum coordinate
conjugate to the phase field $\varphi_{\textbf{r}}$ of
$\Phi^{\dag}_{\textbf{r}}=e^{i\varphi_{\textbf{r}}}$ which is analog
of the position coordinate with commutation relation
$[\varphi_{\textbf{r}},Q_{\textbf{r}}]=i$. The spinon field is
subject to local constraint $|\Phi_{\textbf{r}}|=1$.

We take the continuum limit of this lattice gauge theory by
performing the following expansion to second order in derivative,

\[
\Phi_{\textbf{r}+\widehat{\textbf{e}}_{\mu(\nu)}}=\Phi_{\textbf{r}}+\widehat{\textbf{e}}_{\mu(\nu)}.\nabla
\Phi_{\textbf{r}}+\frac{1}{2}(\widehat{\textbf{e}}_{\mu(\nu)}.\nabla)^2\Phi_{\textbf{r}}+O(\partial^3_{\mu}\Phi)
\]
\begin{equation}\label{phiexpand}
=\Phi_{\textbf{r}}+\partial_{\mu(\nu)}
\Phi_{\textbf{r}}+\frac{1}{2}\partial^2_{\mu(\nu)}\Phi_{\textbf{r}}+O(\partial^3_{\mu(\nu)}\Phi)
\end{equation}

\[
e^{i\int^{\textbf{r}+\widehat{\textbf{e}}_{\mu(\nu)}}_{\textbf{r}}d
\textbf{r}. \textbf{A}
}=e^{i(\widehat{\textbf{e}}_{\mu(\nu)}.\textbf{A}_{\textbf{r},\textbf{r}+\widehat{\textbf{e}}_{\mu(\nu)}})}
\]
\[
e^{iA_{\textbf{r},\textbf{r}+\widehat{\textbf{e}}_{\mu(\nu)}}}=e^{iA_{\mu(\nu)}}
\]
\begin{equation}\label{Aexpand}
=1+iA_{\mu(\nu)}-\frac{1}{2}A^2_{\mu(\nu)}+O(A^3)
\end{equation}

\[
A_{\nu}-A_{\mu}=A_{\textbf{r},\textbf{r}+\widehat{\textbf{e}}_{\nu}}-A_{\textbf{r},\textbf{r}+\widehat{\textbf{e}}_{\mu}}=A_{\textbf{r}+\widehat{\textbf{e}}_{\mu},\textbf{r}+\widehat{\textbf{e}}_{\nu}}
\]
\begin{equation}\label{expansion}
=A_{\textbf{r},\textbf{r}+\widehat{\textbf{e}}_{\nu}-\widehat{\textbf{e}}_{\mu}}=A_{\textbf{r},\textbf{r}+\widehat{\textbf{e}}_{\delta}}=A_{\delta}
\end{equation}

We have defined
$\partial_{\mu(\nu)}=\widehat{\textbf{e}}_{\mu(\nu)}.\nabla$ and
$\widehat{\textbf{e}}_{\delta}=\widehat{\textbf{e}}_{\nu}-\widehat{\textbf{e}}_{\mu}$.
With $\textbf{r},\textbf{r}'$ representing the sites of dual diamond
lattice, the gauge field $A_{\textbf{r},\textbf{r'}}$ lives at the
middle of the link $\frac{\textbf{r}+\textbf{r'}}{2}$ of dual
diamond lattice. At the moment we are essentially working in
Gaussian unit such that we have unit lattice spacing $a=1$, speed of
photon $v_p=1$ and $\hbar=1$, but we will recover these quantities
to their actual physical unit later in the calculation whenever it
is necessary.

One may want to compare the treatment of gauge field fluctuations in
this gauge theory with that in mean field theory where one assumes a
gauge mean field ansatz \cite{3},

\begin{equation}\label{ansatz}
\langle s^{-}_{\mu}\rangle=\frac{1}{2}cos \theta, \langle
s^z_{\mu}\rangle=\frac{1}{2} sin\theta \varepsilon_{\mu}
\end{equation}

where $\varepsilon_{\mu}=(1,1,-1,-1)$ for $\mu=0,1,2,3$
corresponding to the four basis vectors of local cubic base of
pyrochlore lattice in Eq. (\ref{basisvectors}). The QSL state of our
interest was found in gMFT to correspond to $\theta=0$ \cite{3}. We
may equally well write an ansatz,

\begin{equation}
\langle s^{-}_{\mu}\rangle=cos \theta, \langle
s^z_{\mu}\rangle=sin\theta \varepsilon_{\mu}
\end{equation}

Then in the limit
$|A_{\textbf{r},\textbf{r}'}|=|\int^{\textbf{r}'}_{\textbf{r}}\textbf{A}.d\textit{\textbf{r}}|\ll
1$, we have

\[
s^{\pm}_{\textbf{r},\textbf{r'}}=e^{i\pm
A_{\textbf{r},\textbf{r'}}}\simeq (1\pm i
A_{\textbf{r},\textbf{r'}}-\frac{1}{2}A^2_{\textbf{r},\textbf{r'}}+...)
\]

In QSL state, $cos \theta=1$ which gives $\langle s^-\rangle=cos
\theta=1$ and therefore precisely matches with the above expansion
for $s^-_{\textbf{r},\textbf{r}'}$ at lowest order. Physically, this
comparison suggests that the gauge field
$A_{\textbf{r},\textbf{r'}}$ in our expansion is the gauge
fluctuations about the mean field expectation value of gauge field
in gMFT \cite{3}. This is a very accurate physical picture.

The resulting long wavelength theory (still bearing sublattice sums)
is

\begin{widetext}
\[
H=\sum_{\textbf{r}\in I, II}\frac{J_{zz}}{2}Q^2_\textbf{r}
\]
\[
- J_{\pm}\{\sum_{\textbf{r}\in I}\sum_{\mu,\nu\neq\mu}
\{\Phi^{\dag}_{\textbf{r}} \Phi_{\textbf{r}}
+\Phi^{\dag}_{\textbf{r}}(\partial_{\nu}\Phi_{\textbf{r}})+(\partial_{\mu}\Phi^{\dag}_{\textbf{r}})
\Phi_{\textbf{r}}+(\partial_{\mu}\Phi^{\dag}_{\textbf{r}})
(\partial_{\nu}\Phi_{\textbf{r}})+\frac{1}{2}(\partial^2_{\mu}\Phi^{\dag}_{\textbf{r}})
\Phi_{\textbf{r}}+\frac{1}{2}\Phi^{\dag}_{\textbf{r}}
(\partial^2_{\nu}\Phi_{\textbf{r}})+...\}
(1+iA_{\delta}-\frac{1}{2}A_{\delta}^2+...)
\]

\[
- J_{\pm}\{\sum_{\textbf{r}\in II}\sum_{\mu,\nu\neq\mu}
\{\Phi^{\dag}_{\textbf{r}} \Phi_{\textbf{r}}
-\Phi^{\dag}_{\textbf{r}}(\partial_{\nu}\Phi_{\textbf{r}})-(\partial_{\mu}\Phi^{\dag}_{\textbf{r}})
\Phi_{\textbf{r}}+(\partial_{\mu}\Phi^{\dag}_{\textbf{r}})
(\partial_{\nu}\Phi_{\textbf{r}})+\frac{1}{2}(\partial^2_{\mu}\Phi^{\dag}_{\textbf{r}})
\Phi_{\textbf{r}}+\frac{1}{2}\Phi^{\dag}_{\textbf{r}}
(\partial^2_{\nu}\Phi_{\textbf{r}})+...\}
(1-iA_{\delta}-\frac{1}{2}A_{\delta}^2+...)
\]
\[
-J_{z\pm}\{\sum_{\textbf{r}\in
I}\sum_{\mu,\nu\neq\mu}(\Phi^{\dag}_{\textbf{r}}
\Phi_{\textbf{r}}+\Phi^{\dag}_{\textbf{r}}
\partial_{\nu}\Phi_{\textbf{r}}+\frac{1}{2}\Phi^{\dag}_{\textbf{r}}
\partial^2_{\nu}\Phi_{\textbf{r}}+...)E_{\mu}(1+iA_{\nu}-\frac{1}{2}A^2_{\nu}+...)+H.c\}
\]
\begin{equation}\label{latticegaugederivation}
-J_{z\pm}\{\sum_{\textbf{r}\in
II}\sum_{\mu,\nu\neq\mu}(\Phi^{\dag}_{\textbf{r}}
\Phi_{\textbf{r}}-\partial_{\nu}\Phi^{\dag}_{\textbf{r}}
\Phi_{\textbf{r}}+\frac{1}{2}\partial^2_{\nu}\Phi^{\dag}_{\textbf{r}}\Phi_{\textbf{r}}+...)E_{\mu}(1+iA_{\nu}-\frac{1}{2}A^2_{\nu}+...)+H.c\}
\end{equation}
\end{widetext}

where we have denoted
$A_{\mu(\nu)}=A_{\textbf{r},\textbf{r}+\widehat{\textbf{e}}_{\mu(\nu)}}$
in sublattice I and
$A_{\mu(\nu)}=A_{\textbf{r},\textbf{r}-\widehat{\textbf{e}}_{\mu(\nu)}}$
in sublattice II. Likewise,
$E_{\mu(\nu)}=E_{\textbf{r},\textbf{r}+\widehat{\textbf{e}}_{\mu(\nu)}}$
in sublattice I and
$E_{\mu(\nu)}=E_{\textbf{r},\textbf{r}-\widehat{\textbf{e}}_{\mu(\nu)}}$
in sublattice II. The electric field is
$E_{\mu(\nu)}=-\partial_{\mu(\nu)}A_0-\partial_t
A_{\mu(\nu)}=-\partial_t A_{\mu(\nu)}$ since no scalar potential
exists physically in the system. The
$\textbf{r},\textbf{r}'=\textbf{r}\pm
\widehat{\textbf{e}}_{\mu(\nu)}$ here denotes direction (that is,
not the location) of vector to be pointing from $\textbf{r}$ to
$\textbf{r}'=\textbf{r}\pm\widehat{\textbf{e}}_{\mu(\nu)}$. For the
last terms involving coupling $J_{z\pm}$, we will retain only the
lowest order terms in derivatives and gauge fields, valid in the low
energy long distance limit, which must preserve gauge invariance of
the microscopic model Eq. (\ref{latticebosonu1}).

With $\Phi^{\dag}_{\textbf{r}}=e^{i\varphi_{\textbf{r}}}$ and using
commutation relation $[\varphi_{\textbf{r}},Q_{\textbf{r}}]=i$, we
can write $Q_{\textbf{r}}=\frac{1}{i}\frac{\partial}{\partial
\varphi_{\textbf{r}}},\varphi_{\textbf{r}}=\frac{1}{i}\frac{\partial}{\partial
Q_{\textbf{r}}}$. It can be shown that the term $Q^2_{\textbf{r}}$
can be written as
$Q^2_{\textbf{r}}=t^2_s\frac{d\Phi^{*}_{\textbf{r}}}{dt}\frac{d\Phi_{\textbf{r}}}{dt}$
in the field theory language where $t_s$ is an appropriate time
scale needed to get the dimension right. We choose this time scale
to be $t_s=\frac{\hbar}{J_{zz}}$($\equiv\frac{1}{J_{zz}}$ in
Gaussian unit) because the coupling $J_{zz}$ is the reference energy
scale (coupling constant) in the phase diagram. One may want to
minimally couple this term to the the scalar potential $A_0$ to get
gauge invariant term
$Q^2_{\textbf{r}}=|(i\partial_{t}-A_0)\Phi_{\textbf{r}}|^2$ but
physically, there exists no scalar potential in microscopic model in
Eq. (\ref{latticebosonu1}), i.e. $A_0=0$ and only vector potential
$\textbf{A}$ exists with the electric field $\textbf{E}$ coming
entirely from this vector potential.

We obtain from Eq. (\ref{latticegaugederivation}) a $3+1$-D
continuum action in real (Minkowskian) time ($T=0$) field theory
\cite{psqft},

\begin{widetext}
\[
S=\int d^4 x
[\frac{1}{2J_{zz}}|(i\partial_{t}-e_gA_0)\Phi_{\textbf{r}}|^2-
\frac{1}{2}J_{\pm}\sum_{\mu,\nu\neq\mu}|(i\partial_{\delta}-eA_{\delta})\Phi_{\textbf{r}}|^2
-(\lambda-12J_{\pm})|\Phi_{\textbf{r}}|^2-\frac{1}{2g^2}\sum_{\alpha\beta}(\partial_{\alpha}A_{\beta}-\partial_{\beta}A_{\alpha})^2
\]
\begin{equation}\label{effaction}
+\frac{1}{2g^2}\sum_{\alpha}E^2_{\alpha}+\frac{1}{2}J_{z\pm}\{\sum_{\mu,\nu\neq\mu}\gamma^*_{\mu\nu}ieE_{\mu}J_{\nu}+H.c.\}-\frac{1}{2}J_{z\pm}\sum_{\mu,\nu\neq\mu}\{\gamma^*_{\mu\nu}eE_{\mu}|(i\partial_{\nu}-eA_{\nu})\Phi_{\textbf{r}}|^2+H.c.\}]
\end{equation}
\end{widetext}

with the gauge invariant Noether current $J_{\nu}$ given by

\begin{equation}\label{current}
J_{\nu}=\Phi^*_{\textbf{r}}(-i\partial_{\nu}+eA_{\nu})\Phi_{\textbf{r}}+\Phi_{\textbf{r}}(i\partial_{\nu}+eA_{\nu})\Phi^*_{\textbf{r}}
\end{equation}

and the partition function

\[
Z=\int \mathcal{D} \Phi^{*} \mathcal{D} \Phi \mathcal{D}A
\mathcal{D}\lambda e^{iS[\Phi^{*},\Phi,A,\lambda]}
\]

where we have added the Lagrange multiplier term $\lambda\int d^4
x(\Phi^{*}_{\textbf{r}}\Phi_{\textbf{r}}-1)$ with $\lambda>0$ that
globally imposes the Hilbert space constraint on the spinon field
$|\Phi_{\textbf{r}}|=1$. The $\lambda\int d^4 x (-1)$ piece
contributes only a constant energy shift and is omitted from Eq.
(\ref{effaction}). The Hermitian conjugation operation ($^\dagger$)
in the original Hamiltonian language becomes simply complex
conjugation operation ($^*$) in the field theory language as the
original bosonic spinon creation and annihilation operators now
simply become complex scalar field.

The $\mu,\nu=0,1,2,3$ are the indices of local $z$ spin axes
$\widehat{\textbf{e}}_{\mu(\nu)}$ and $\alpha,\beta=x,y,z$. The mass
parameter $m=\lambda-12J_{\pm}$ is the spinon gap. We have put in by
hand the free Maxwell term of Abelian $U(1)$ gauge theory, separated
into its magnetic field and electric field parts, analogous to the
Maxwell term of QED: $-\frac{1}{4}F_{\mu\nu}F^{\mu\nu}$ in Gaussian
unit, with $\mu=g^2\mu_0$. This free Maxwell action corresponds to
$-\frac{1}{2}(c^2\textbf{B}^2-\textbf{E}^2)$ which is standard in
$U(1)$ gauge theory and also generally describes very well the
actual physics of Coulomb phases in pyrochlore, with certain
cautions. We have also included the emergent $U(1)$ gauge charge
$e_g\equiv Q$ (or equivalently $e$ which we define as
$e=\frac{4}{3}e_g$) together with each of the gauge fields $A$ and
$E$ which represents the strength of spinon-gauge field coupling.

The field theory in Eq. (\ref{effaction}), apart from the last two
terms, is called scalar QED in QFT. This scalar QED part is truly
Lorentz invariant when the coefficients of the first two terms are
equal. The last two terms are novel terms that reflect the unique
physics of field theory of pyrochlore QAFM, derived directly from
the microscopic model Eq. (\ref{latticebosonu1}). We treat these
last two terms as perturbation to the scalar QED part. This is
justified by the fact that these last two terms have coupling
proportional to $U(1)$ gauge charge $e$ (or equivalently $e_g$) and
because these terms are linear in electric field, they should be
multiplied with an inverse of mass scale in order to have proper
mass dimension. This mass scale is nothing but a UV cut off
$\Lambda$; a large momentum (mass) scale which we can take as the
inverse of the small lattice spacing $a$. Further, these last two
terms give rise to several new types of vertex; the simplest ones
being scalar-scalar-gauge field vertex and scalar-scalar-gauge
field-gauge field vertex. However, we will not discuss the
renormalization effect of these terms or explicitly compute their
contribution to the renormalization correction of the appropriate
terms of the scalar QED but only give the order of magnitude of
those corrections. The two most important renormalization effects of
those vertices are mass renormalization and quartic term
renormalization \cite{quarticJpmpm}. Denoting them as $\delta
m^{z\pm}$ and $\delta u^{z\pm}$ respectively, it is easy to check
that the leading contributions are $\delta
m^{z\pm}=\mathcal{O}(e^2J^2_{z\pm})$ and $\delta
u^{z\pm}=\mathcal{O}(e^4J^2_{z\pm})$.

The new field theory upon taking the above consideration now becomes

\begin{widetext}
\begin{equation}\label{effactioncopy}
S=\int d^4 x [\frac{1}{2J_{zz}}|(i\partial_{t}-e_g
A_0)\Phi_{\textbf{r}}|^2-
\frac{16}{3}J_{\pm}\sum_{\alpha}|(i\partial_{\alpha}-e_g
A_{\alpha})\Phi_{\textbf{r}}|^2
-m|\Phi_{\textbf{r}}|^2-u|\Phi_{\textbf{r}}|^4-\frac{1}{2g^2}\sum_{\alpha\beta}(\partial_{\alpha}A_{\beta}-\partial_{\beta}A_{\alpha})^2+\frac{1}{2g^2}\sum_{\alpha}E^2_{\alpha}]
\end{equation}
\end{widetext}

where we have used the mapping Eq. (\ref{basisvectorsmapping}) and
the gauge field mapping \cite{Amapping} to express all the vectors
in global Cartesian coordinate basis. We have also obtained the
spinon gap $m=\lambda-12J_{\pm}$.

\bigskip

\section{Free Energy Description of QSL-AFM QPT }

In this section, we give the details of the derivation of (quantum
analog of the classical thermal) "free energy" for bosonic spinon
fields to be used to describe QSL-AFM quantum phase transition where
the expectation value of spinon field $\langle\Phi\rangle$ is the
order parameter for this transition. The field theory for pyrochlore
QAFM in imaginary time (Euclidean space-time) \cite{7} with
$t=-i\tau$ is described by

\begin{widetext}
\begin{equation}\label{effactionEuc}
S^E=\int d^4 x^E [\frac{1}{2J_{zz}}|(\partial_{\tau}+ie_g
A_0)\Phi_{\textbf{r}}|^2+\frac{16}{3}J_{\pm}\sum_{\alpha}|(\partial_{\alpha}+ie_g
A_{\alpha})\Phi_{\textbf{r}}|^2
+m|\Phi_{\textbf{r}}|^2+u|\Phi_{\textbf{r}}|^4+\frac{1}{2g^2}\sum_{\alpha\beta}(\partial_{\alpha}A_{\beta}-\partial_{\beta}A_{\alpha})^2+\frac{1}{2g^2}\sum_{\alpha}E^2_{\alpha}]
\end{equation}
\end{widetext}

We aim for free energy $F[\Phi^*,\Phi]$ to describe the QSL-AFM
phase transition because such transition is based on bosonic spinon
condensation where $\langle \Phi\rangle=0$ in QSL and
$\langle\Phi\rangle\neq 0$ in AFM. It is to be noted in both of
these phases, $\langle E\rangle=0$ \cite{3}. From Eq.
 (\ref{effactionEuc}), the real space free energy density of spinon
in the static spatially uniform approximation \cite{13}, obtained by
taking $\partial_{\alpha}\Phi_{\textbf{r}}=0$, gives

\[
\frac{1}{V}\frac{d
F[\Phi^*_{\textbf{r}},\Phi_{\textbf{r}}]}{d|\Phi_{\textbf{r}}|}=2m
|\Phi_{\textbf{r}}|+4u
|\Phi_{\textbf{r}}|^3+J_{\pm}e^2\sum_{\mu,\nu\neq\mu} \langle
A^2_{\delta(\textbf{r})}\rangle_{\Phi}|\Phi_{\textbf{r}}|
\]
\begin{equation}\label{effactionPhi}
=2m |\Phi_{\textbf{r}}|+4u
|\Phi_{\textbf{r}}|^3+6J_{\pm}e^2\sum_{\alpha\beta} \langle
A_{\alpha(\textbf{r})}A_{\beta(\textbf{r})}\rangle_{\Phi}\delta_{\alpha\beta}|\Phi_{\textbf{r}}|
\end{equation}

where $V$ is the system volume. In the action language in Eq.
(\ref{effactionEuc}), each $\int_k$ is $4$-d momentum-frequency
integral $\int \frac{d\omega}{2\pi}\int\frac{d^3k}{(2\pi)^3}$
whereas in this free energy language, each $\int_{k}$ is $3$-d
momentum integral $\int\frac{d^3k}{(2\pi)^3}$ with the integration
over frequency turns into prefactor proportional to inverse of an
energy scale, as shown later.

The expectation value $\langle...\rangle$ in Eq.
(\ref{effactionPhi}) is evaluated with respect to appropriate
action. The next step is therefore to compute these expectation
values. Taking Fourier transform, we have to compute $\langle
A_{\delta(k'')}A_{\delta(k''')}\rangle_{\Phi}$ which is the
expectation value of $A_{\delta(k'')}A_{\delta(k''')}$ with respect
to the action (free energy) of $A$ taken at constant spatially
uniform value of $\Phi$.

\[
\langle A_{\delta(k'')}A_{\delta(k''')}\rangle_{\Phi}=\frac{\int
\mathcal{D}\textbf{A}
e^{-\frac{F[\textbf{A}]_{\Phi}}{T_{QPT}}}A_{\delta(k'')}A_{\delta(k''')}}{\int
\mathcal{D}\textbf{A}e^{-\frac{F[\textbf{A}]_{\Phi}}{T_{QPT}}}}
\]
\[
=\sum_{\alpha\beta}(\frac{3}{4})^2n_{\delta\alpha}n_{\delta\beta}\frac{\int
\mathcal{D}\textbf{A}
e^{-\frac{F[\textbf{A}]_{\Phi}}{T_{QPT}}}A_{\alpha(k'')}A_{\beta(k''')}}{\int
\mathcal{D}\textbf{A}e^{-\frac{F[\textbf{A}]_{\Phi}}{T_{QPT}}}}
\]
\[
=(\frac{3}{4})^2\sum_{\alpha\beta}n_{\delta\alpha}n_{\delta\beta}\langle
A_{\alpha(k'')}A_{\beta(k''')}\rangle_{\Phi}
\]

where $J_{\pm}$ plays the role of energy scale $T_{QPT}$ that tunes
this quantum phase transition. The free energy of the gauge field is

\begin{widetext}
\[
F[\textbf{A}]_{\Phi}=\frac{1}{2g^2}\int_{k'',k'''}
A_{\alpha(k'')}(k''^2\delta_{\alpha\beta}-k''_{\alpha}k''_{\beta})A_{\beta(k''')}\delta(k''+k''')+\frac{1}{2}J_{\pm}\sum_{\mu,\nu\neq\mu}\int_{k,k',k'',k'''}e^2A_{\delta(k'')}A_{\delta(k''')}\langle\Phi^*_k\Phi_{k'}\rangle\delta(-k+k'+k''+k''')
\]
\begin{equation}
=\frac{1}{2g^2}\int_{k'',k'''}
A_{\alpha(k'')}(k''^2\delta_{\alpha\beta}-k''_{\alpha}k''_{\beta})A_{\beta(k''')}\delta(k''+k''')+3J_{\pm}\sum_{\alpha\beta}\int_{k,k',k'',k'''}e^2A_{\alpha(k'')}A_{\beta(k''')}\langle\Phi^*_k\Phi_{k'}\rangle\delta(-k+k'+k''+k''')
\end{equation}
\end{widetext}

where we have used

\[
\sum_{\mu,\nu\neq\mu}A_{\delta(k'')}A_{\delta(k''')}
=(\frac{3}{4})^2\sum_{\mu,\nu\neq\mu}\sum_{\alpha\beta}n_{\delta\alpha}A_{\alpha(k'')}n_{\delta\beta}A_{\beta(k''')}
\]
\begin{equation}
=6\sum_{\alpha\beta}A_{\alpha(k'')}A_{\beta(k''')}\delta_{\alpha\beta}
\end{equation}

Noting that $\langle\Phi^*_{k}\Phi_{k'}\rangle\sim \delta(k'-k)$ and
using

\[Z_{\textbf{A}}=\int \mathcal{D}\textbf{A}
e^{-\int_{k'',k'''}A_{\alpha(k'')}M^{\Phi}_{\alpha\beta}(k'',k''')A_{\beta(k''')}\delta(k''+k''')}
\]
\begin{equation}
=\int
\mathcal{D}\textbf{A}e^{-S[\textbf{A}]}=\frac{1}{M^{\Phi}_{\alpha\beta}}\delta(k''+k''')
\end{equation}

and

\[\langle A_{\alpha(k'')}A_{\beta(k''')} \rangle=\frac{\int
\mathcal{D}\textbf{A}
e^{-S[\textbf{A}]}A_{\alpha(k'')}A_{\beta(k''')}}{\int
\mathcal{D}\textbf{A} e^{-S[\textbf{A}]}}
\]
\begin{equation}
=\frac{-\frac{\partial Z_{\textbf{A}}}{\partial
M^{\Phi}_{\alpha\beta}}}{Z_{\textbf{A}}}=\frac{1}{M^{\Phi}_{\alpha\beta}}\delta(k''+k''')
\end{equation}

, the result is

\begin{equation}\label{AAcorr1}
\langle A_{\alpha(k'')}A_{\beta(k''')}\rangle_{\Phi}
=\frac{2g^2J^c_{\pm}}{k''^2+6J_{\pm}e^2g^2|\Phi|^2}(\delta_{\alpha\beta}-\frac{k''_{\alpha}k''_{\beta}}{\textbf{k''}^2})\delta_{k'',-k'''}
\end{equation} and so

\[
\langle A_{\delta(k'')}A_{\delta(k''')}\rangle_{\Phi}=
\]
\begin{equation}\label{AAcorr2}
(\frac{3}{4})^2\sum_{\alpha\beta}\frac{2g^2J^c_{\pm}n_{\delta\alpha}n_{\delta\beta}}{k''^2+6J_{\pm}e^2g^2|\Phi|^2}(\delta_{\alpha\beta}-\frac{k''_{\alpha}k''_{\beta}}{\textbf{k''}^2})\delta_{k'',-k'''}
\end{equation}

where $J^c_{\pm}=\frac{\lambda}{12}$ is the critical $J_{\pm}$ at
which $m$ changes sign. Retaining only the gauge independent part of
the transverse projector, as only this part that should contribute
to physical process, we have

\begin{widetext}
\begin{equation}
\frac{1}{V}\frac{d
F[\Phi^*_{\textbf{r}},\Phi_{\textbf{r}}]}{d|\Phi_{\textbf{r}}|}=2m
|\Phi_{\textbf{r}}|+4u
|\Phi_{\textbf{r}}|^3+6J_{\pm}e^2\sum_{\alpha\beta}\int_{k''}\frac{2g^2J^c_{\pm}}{k''^2+6J_{\pm}e^2g^2|\Phi|^2}\delta_{\alpha\beta}|\Phi_{\textbf{r}}|
\end{equation}
\end{widetext}

where we have used
$\sum_{\mu,\nu\neq\mu}n_{\delta\alpha}n_{\delta\beta}=\frac{32}{3}\delta_{\alpha\beta}$
to obtain the last line. The integral $\int_{
k''}\frac{2g^2J^c_{\pm}}{k''^2+6J_{\pm}e^2g^2|\Phi|^2}\delta_{\alpha\beta}$
is UV divergent and so we impose UV cut-off. Denoting
$k^2_s=6J_{\pm}e^2g^2|\Phi|^2$, we obtain

\[
\int
\frac{d^3k''}{(2\pi)^3}\frac{1}{k''^2+k^2_s}=
\frac{1}{(2\pi)^3}4\pi(\Lambda
- k_s tan^{-1}[\frac{\Lambda}{k_s}])
\]
\begin{equation}\label{integralAA}
\simeq \frac{1}{2\pi^2}(\Lambda - k_s \frac{\pi}{2})
\end{equation}

in the limit of large $\frac{\Lambda}{k_s}$. To obtain the final
free energy density in real space, we recover the spatial dependence
of the $\Phi$ and impose locality of the free energy and get

\begin{widetext}
\begin{equation}
\frac{1}{V}\frac{d
F[\Phi^*_{\textbf{r}},\Phi_{\textbf{r}}]}{d|\Phi_{\textbf{r}}|}=2m
|\Phi_{\textbf{r}}|+4u
|\Phi_{\textbf{r}}|^3+\frac{6e^2g^2}{\pi^2}J_{\pm}J^c_{\pm}(\Lambda-\frac{\pi}{2}\sqrt{6J_{\pm}e^2g^2|\Phi_{\textbf{r}}|^2})\sum_{\alpha\beta}\delta_{\alpha\beta}|\Phi_{\textbf{r}}|
\end{equation}
\[
=2c_2|\Phi_{\textbf{r}}|-3c_3|\Phi_{\textbf{r}}|^2+4u|\Phi_{\textbf{r}}|^3
\]
\end{widetext}

where $c_2=m+\delta m$ with $m=\lambda-12J_{\pm},\delta
m=\frac{9J_{\pm}e^2g^2J^c_{\pm}\Lambda}{\pi^2}+\mathcal{O}(e^2J^2_{z\pm})$,
$c_3=\frac{3J_{\pm}e^2g^2J^c_{\pm}}{\pi}\sqrt{6J_{\pm}e^2g^2}$ in
Gaussian unit, and $\Lambda\sim \frac{1}{a}$ with $a$ is microscopic
lattice spacing. In physical unit, $\delta
m=\frac{9J_{\pm}a^2e^2g^2\mu_0J^c_{\pm}\Lambda}{\pi^2\hbar^2}+\mathcal{O}(J^2_{z\pm}e^2)$,
$c_3=\frac{3J_{\pm}a^2e^2g^2\mu_0J^c_{\pm}}{\pi\hbar^2}\sqrt{\frac{6J_{\pm}a^2e^2g^2\mu_0}{\hbar^2}}$,
$c_4\equiv u=u_0+\mathcal{O}(J^2_{z\pm}e^4)$ \cite{quarticJpmpm}
where we have used the result in Appendix A for the last correction
to $\delta m$ and $u$.

The final free energy in real space takes the form,

\begin{equation}\label{finalfreeenergy}
F[\Phi^*_{\textbf{r}},\Phi_{\textbf{r}}]=\int d^3 r
[c_2|\Phi_\textbf{r}|^2-c_3|\Phi_\textbf{r}|^3+c_4|\Phi_\textbf{r}|^4]
\end{equation}

with
$c_2=\lambda-12J_{\pm}+\frac{16J_{\pm}a^2e^2_gg^2\mu_0J^c_{\pm}\Lambda}{\pi^2\hbar^2}+\mathcal{O}(e^2_gJ^2_{z\pm})$,
$c_3=\frac{16J_{\pm}a^2e^2_g
g^2\mu_0J^c_{\pm}}{3\pi\hbar^2}\sqrt{\frac{32J_{\pm}a^2e^2_g
g^2\mu_0}{3\hbar^2}}$, $c_4\equiv
u=u_0+\mathcal{O}(J^2_{z\pm}e^4_g)$ where we have used
$e_g=\frac{3}{4}e$.

The location of QSL-AFM phase transition can be predicted directly
from the free energy Eq. (\ref{finalfreeenergy}). We noticed
previously that the coupling to gauge fields renormalizes the spinon
gap (mass) $m$ only by subleading correction $\delta
m=\frac{16J_{\pm}a^2e^2_gg^2\mu_0J^c_{\pm}\Lambda}{\pi^2\hbar^2}+\mathcal{O}(e^2_gJ^2_{z\pm})$
(in physical unit). According to Eq. (\ref{finalfreeenergy}), it can
be shown that the phase transition occurs at

\begin{equation}\label{transpoint}
c_2=\frac{c^2_3}{4c_4}
\end{equation}

Physically, the QSL to AFM phase transition is bosonic spinon
condensation that occurs once the spinon becomes gapless. With $c_3$
and $c_4$ given as before, Eq. (\ref{transpoint}) suggests that the
transition occurs at $m=\frac{c^2_3}{4c_4}-\delta m$. Since both
terms on the right hand side are subleading to $m$, to lowest order
approximation, the QSL-AFM phase transition therefore occurs at
$m=\lambda-12J_{\pm}=0$ or equivalently $\lambda=12J_{\pm}$. To
compare this with gMFT result however, we have to carefully take
into account an extra factor of $\frac{1}{4}=(\frac{1}{2})^2$ which
arises from the fact that the gMFT ansatz \cite{3} Eq.
(\ref{ansatz})

\begin{equation}
\langle s^{-}_{\mu}\rangle=\frac{1}{2}cos \theta, \langle
s^z_{\mu}\rangle=\frac{1}{2} sin\theta \varepsilon_{\mu}
\end{equation}

matches precisely with the spin-gauge field correspondence Eq.
(\ref{correspondence})

\begin{equation}
s^z_{\textbf{r},\textbf{r'}}=E_{\textbf{r},\textbf{r'}},s^{\pm}=e^{i\pm
A_{\textbf{r},\textbf{r'}}}
\end{equation}

only if we add factor half to the right hand side of the
correspondence for $s^{\pm}$, i.e.,
\[
s^{\pm}=\frac{1}{2}e^{i\pm A_{\textbf{r},\textbf{r'}}}
\]
Therefore, since the $J_{\pm}$ term in the lattice gauge theory Eq.
(\ref{latticebosonu1}) consists of products of bilinear term in
bosonic spinon fields $\Phi^{\dag},\Phi$ and bilinear term in
(exponential of gauge fields) $s^{+},s^{-}$, it effectively predicts
QSL-AFM phase transition at
$\lambda=(\frac{1}{2})^212J_{\pm}=3J_{\pm}$, in precise agreement
with Ref. \cite{3}.

The size of first order phase transition can be obtained by taking
$F[|\Phi|]=0,\frac{\partial F[|\Phi|]}{\partial|\Phi|}=0$ from Eq.
 (\ref{finalfreeenergy}) which gives $\Delta
\langle\Phi\rangle=\frac{c_3}{2c_4}|_{QPT}$. If we measure the
strength of first order phase transition by the ratio of the jump
$\Delta \langle\Phi\rangle$ to the magnitude of the order parameter
$\langle\Phi\rangle_0$ deep inside the magnetically ordered AFM
state, derivable from Eq. (\ref{finalfreeenergy}) by taking
$\frac{\partial F[|\Phi|]}{\partial|\Phi|}=0$ and the limit
$J_{z\pm}\ll J_{\pm}\sim J_{zz}$,
$\langle\Phi\rangle_0\simeq\frac{3c_3}{4c_4}|_{AFM}$, we obtain
$\frac{\Delta
\langle\Phi\rangle}{\langle\Phi\rangle_0}=\frac{2}{3}(\frac{J^c_{\pm}}{J_{zz}})^{\frac{3}{2}}\comment{\frac{\langle
A_{\alpha}A_{\beta}\rangle^2|_{AFM}}{\langle
A_{\alpha}A_{\beta}\rangle^2|_{QPT}}\simeq}\lesssim 0.06$ using
$J^c_{\pm}\simeq 0.2J_{zz}$, which is roughly the location of
QSL-AFM phase transition predicted by gMFT \cite{3}. This result
therefore suggests a relatively weak first order phase transition.

As a final note, so far we have not specified explicitly the value
of parameters in the field theory Eq. (\ref{effaction}) such as the
effective $U(1)$ gauge charge $e$ (or $e_g$) and permeability ratio
$g^2=\frac{\mu}{\mu_0}$ whereas other parameters  such as lattice
spacing $a$ should be measurable and known for each specific
compound. These parameters characterize the emergent electrodynamics
and should be treated as phenomenological quantities determinable
from experiment.


\begin{thebibliography}{1}
\bibitem{gaugecharge}Physically, $Q=\pm 1$ in the lattice gauge theory
Eq. (\ref{latticegauge}) but the corresponding charge $e_g\sim Q$ is
small parameter in field theory Eq. (\ref{effaction1}) as is the
case in standard QED \cite{psqft}.
\bibitem{loopcorrection}The last correction to $\delta m$ given by
its order of magnitude $\mathcal{O}(e^2_gJ^2_{z\pm})$ and the
correction to quartic coefficient $\mathcal{O}(e^4_g J^2_{z\pm})$
come from novel terms with coupling constant proportional to
$J_{z\pm}$ which we treat as perturbation and give rise to loop
corrections to the appropriate terms in the scalar QED; quadratic
(mass) and quartic terms respectively in the field theory Eq.
(\ref{effaction1}). Please see discussion at the latter part of
Appendix A for the details on this. We do not explicitly compute
these loop corrections and defer them for future publication. These
loop corrections also do not change the final conclusions of this
work qualitatively.
\bibitem{quarticJpmpm} Quartic term is needed from the beginning to ensure both the stability and the renormalizability of the field theory.
Microscopically in the context of this work, it can be provided by
the $J_{\pm\pm}[\gamma_{ij}S^+_iS^+_j+\gamma^*_{ij}S^-_iS^-_j]$ term
in the microscopic spin model \cite{3}, where in this case $u_0\sim
J_{\pm\pm}$. But including such term will give rise to a slightly
different mean field phase diagram with $U(1)$ QSL and
antiferromagnetic quadrupolar (AFQ) order \cite{SBLee} compared to
AFM for the case with $J_{\pm\pm}=0$ \cite{3}. We may however
consider the case with very small $J_{\pm\pm}\ll J_{\pm},
J_{z\pm}\lesssim J_{zz}$ which is sufficient to ensure the stability
of the field theory describing the phase transition and yet can
still be used to describe phase diagram with $J_{\pm\pm}=0$
 \cite{3}. From more general perspective, spin exchange model of the
form $J^{ab}_{ij}S^a_{i}S^b_{j}$ followed by slave-particle type
decomposition
$S^{a(b)}_{i}=\frac{1}{2}\Phi^{\dag}_{i\alpha}\sigma^{a(b)}_{\alpha\beta}\Phi_{i\beta}$
will in general generate quartic term from the product of two spin
operators which gives rise to product of four spinon operators and
in this case $u_0\sim J^{ab}_{ij}$ as a result.
\end{thebibliography}
\end{document}